\documentclass[%
 reprint,
superscriptaddress,
 amsmath,amssymb,
 aps,
]{revtex4-2}

\usepackage{graphicx}
\usepackage{dcolumn}
\usepackage{bm}
\usepackage{hyperref}

\hypersetup{
    colorlinks,
    citecolor=blue,
    filecolor=black,
    linkcolor=black,
    urlcolor=black
}

\begin{document}

\preprint{APS/123-QED}

\title{Modelling Micropipette Aspiration with Active Particles}

\author{G. Ourique}
\affiliation{Universidade Federal do Rio Grande do Sul}

\author{E. F. Teixeira}%
\affiliation{Universidade Federal do Rio Grande do Sul}

\author{L. G. Brunnet}%
\affiliation{Universidade Federal do Rio Grande do Sul}

\date{\today}

\begin{abstract}

The study of cells' dynamical properties is essential to a better understanding of several physiological processes. These properties are directly associated with cells' mechanical parameters experimentally achieved through physical stress. The micropipette aspiration essay has proven an accurate and controllable tool to apply physical stress to the cell. In this work, we explore the numerical modeling of two-dimensional cells using an active multi-particle ring submitted to micropipette aspiration. We correlate simulation parameters with experimental data and obtain a complete map of the input parameters and the resulting elastic parameters that could be measured in experiments.

\end{abstract}

\maketitle

\section{\label{sec:intro}Introduction}

Cell migration is essential in physiological processes such as embryogenesis, wound healing, and tumor cells' metastasis~\citep{Haga2005}. It is directly associated with cell mechanical behavior~\citep{Houk2012, VanHelvert2018}, motivating the development of several experiments to quantify forces ruling individual cell dynamics~\citep{Rodriguez2013}.
Experiments determining the mechanical properties of cells use well-defined forces to deform cells and measure a deformation response~\citep{Hochmuth1993}. In particular, micropipette aspiration experiments provide time-dependent analysis of living cells~\citep{Hochmuth2000,Rodriguez2013} or cell aggregates~\citep{Guevorkian2010, Guevorkian2017}, being useful to determine cells' viscoelastic properties~\citep{Yeung1989,Athanasiou1999, Jones1999}.

 Mechanical properties of individual cells can be described by simple analytical models~\citep{Theret1988,Wu1999, Hochmuth2000,Plaza2015}. However, these models are limited to stationary solutions of symmetrical membranes, not reproducing  heterogeneity, anisotropy and dynamical changes observed in real cells~\citep{Gonzalez-Bermudez2019}.

Numerical simulation modeling may include cell structure details. For example, Monte Carlo simulations can describe non-homogeneous and anisotropic properties of membrane structures~\citep{Boey1998,Discher1998}. More sophisticated numerical simulations can describe cell membrane, nucleus, and cytoskeleton using dynamical evolution of several dissipative particles connected~\citep{Lykov2017}, recovering elastic properties from cells observed in experiments.

One successful model to describe flexible bodies such as cells is a ``spring-bead-like'' model, consisting of several particles (beads) connected under a harmonic potential in a closed-ring structure. Due to its simplicity, the simulations may easily include features like cell duplication. \citet{Teixeira2021} introduced such a model with particles replaced by overdamped active particles, similar to the individual particles presented by \citet{Szabo2009}. This approach makes the entire structure move as a complex and flexible active particle, being the right candidate for a minimalist cell representation model.

However, the model presented by \citet{Teixeira2021} is not suitable to reproduce the micropipette aspiration experiment. Due to the lack of internal structure, we may compress the ring to the limit where its opposite sides touch. 
Therefore, above some pressure difference, cells ruled by this model would always get totally inside the micropipette when aspirated. In this work, we propose a modified version of this model, including cell volume conservation. We explore the relationship between the externally applied pressure and the cell deformation length inside the micropipette. Moreover, we compare the numerical results with the theoretical description made by \citet{Theret1988}.

This paper is organised as follows: Section \ref{sec:modeldesc} describes the extension of \citet{Teixeira2021} model used in this work. Section \ref{sec:results} describes simulation results, and in Section \ref{sec:conclusion} we conclude.

\section{\label{sec:modeldesc}Model Description}

\citet{Teixeira2021} model represents a two-dimensional cell membrane using several particles connected, forming a ring-like structure maintained by three potentials: a harmonic bond potential, which keeps particle distances around an equilibrium value; a bending potential, which maintains the angle between neighbouring particles close to zero;  and a pure repulsive Weeks-Chandler-Andersen (WCA) potential~\citep{Weeks1971} representing excluded-volume interaction between particles.
Our modified version simplifies the forces derived from the  WCA potential to purely linear repulsive, providing less stiff particle-particle interactions in the membrane. We also include a potential term to preserve cell area around a target one. 

\subsection{Membrane Description}
We construct the cell membrane using N connected particles in a ring-like structure.
 The harmonic potential $U_s$, between neighbouring  particles is,
\begin{equation} \label{harmonic_potential}
    U_s {=} \frac{k_s}{2}\left(\sum_{k=1}^{N-1}\left(|\vec{d}_{k,k+1}| {-} d_0\right)^2 +\left(|\vec{d}_{N,1}| {-} d_0\right)^2\right)
\end{equation}
where $k_s$  defines the spring rigidity, $\vec{d}_{i,j}$ is the distance between $i$-th and $j$-th membrane particles, i.e. $\vec{d}_{i,j}=\vec{r}_{j}-\vec{r}_{i}$, $d_0$ is the equilibrium distance. The last term in Eq. \ref{harmonic_potential} relates the last ($N$) and the first particles, configuring a closed-ring structure for the membrane.

A bending potential, $U_b$, controls the angle between between neighbouring particles,
\begin{equation}\label{eq:potentialbend}\small{
    U_b {=} \frac{k_b}{2}\left(N {-} \sum_{k=1}^{N-1}\cos(\theta_{k,k+1} {-} \theta_0) {+} \cos(\theta_{N,1} {-} \theta_0)\right)
    }
\end{equation}
where $k_b$ is the bending rigidity, $\theta_{i,j}$ is the angle between vectors $\vec{d}_{i,j}$ and $\vec{d}_{i+1,j+1}$, and $\theta_0$ is the equilibrium angle.

To treat any two-particle interactions non-consecutive in the ring, we define a linear repulsion force, as \citet{Szabo2006} proposed.
The potential related to this force, $U_r$, is given by,

\begin{equation}\label{eq:purerepulsive} \small{
    U_r = \sum_{i=1}^{N-1} \sum_{j=i+1}^N \left\{\begin{array}{ll}
                \frac{k_r |\vec{d}_{i,j}|}{2 d_e}\left(|\vec{d}_{i,j}|{-}2 d_e\right), &\, |\vec{d}_{i,j}| {\leq} d_e\\
                 -\frac{k_r d_e}{2} , &\, |\vec{d}_{i,j}| {>} d_e
                \end{array}\right.
                }
\end{equation}
where $k_r$ is the maximum amplitude of the repulsive force and $d_e$ is the maximum distance of the interaction.

\subsection{Cytoplasm and Pipette Wall Description}

Since the cytoplasm is  nearly incompressible  \cite{Hartono2011}, we model it using a harmonic potential, $U_A$, around a target area surrounded by the membrane, \begin{equation}\label{eq:potentialarea}
    U_A = \frac{k_A}{2}\left(A - A_0\right)^2
\end{equation}
where $k_A$ is the harmonic potential constant, which defines the compressibility of the cytoplasm, $A$ is the area inside the membrane perimeter, and $A_0$ is the equilibrium area. Since we are working in a two-dimensional model, the area plays the equivalent role of the volume in three-dimensions.

This harmonic potential implies that cell volume will be conserved, allowing small fluctuations ruled by the harmonic potential constant. The force associated with this potential pulls and pushes the membrane perpendicularly, acting as a cortical tension.

We describe the wall as a sequence of small repulsive particles, tracing the wall form. The spacing of the particles that compose the wall is $d_0/2$, ensuring that any particle from the membrane can pass through the wall and the total repulsive force applied to the membrane is smoothly defined along side the micropipette channel. The sharpness of the micropipette edges were smoothed to make the simulated micropipette closer to the real micropipettes.

\subsection{Dynamic Equations}

We adopted the same overdamped self-propelled dynamics described by \citet{Szabo2006} for the simulations.  Each particle of the system presents a self-propulsion oriented in a polarization direction, which gradually aligns with the resultant force applied to the particle. As already shown by \citet{Szabo2006}, after appropriate parameter tuning, this dynamics results in a collective motion without the need of averaging neighbouring particles velocities, resulting in a suitable approach for cell movement description. Equations of motion for particle $i$ are 
\begin{eqnarray}
\frac{d \vec{r}_i (t)}{dt} &=& v_0 \vec{n}_i(t) + \mu\vec{F}_i(t) \label{eq:drdt} \\
\frac{d\theta_i(t)}{dt} &=& \frac{1}{\tau}  \arcsin{\left(\left(\vec{n}_i(t)\times \frac{\vec{v}_i(t)}{|\vec{v}_i(t)|} \right)\cdot \vec{e}_z \right)} + \xi_i(t) \label{eq:dthetadt}
\end{eqnarray}
where $v_0$ is the self-propulsion velocity, $\theta_i$ is the particle orientation, $\vec{n}_i=(\cos{(\theta_i)}\hat{x}+\sin{(\theta_i)}\hat{y})$, $\mu$ is the mobility, $\vec{F}$, derived from the previously defined potentials, is the resultant force over the particle, $\tau$ is the relaxation time and $\xi_i$ is a zero-mean white noise with standard deviation equals to $\sqrt{2D_R}$. $D_R$ is the angular noise coefficient, which is the inverse of the persistence time, $\tau_r$. We use a low noise level in the simulations,  $\tau_r = 100\tau$.

\citet{Teixeira2021} demonstrated in their work that a ring of active particles connected by spring-like forces following the dynamics proposed by \citet{Szabo2006} acts like an extended active body. Under appropriate parameters, connections force each particle's orientation alignment, inducing the ring to collective motion. For the 2d single ring, two main collective motions are possible, rotation or translation, depending only on initial conditions.

\begin{figure}[htp]
\includegraphics[width=\linewidth]{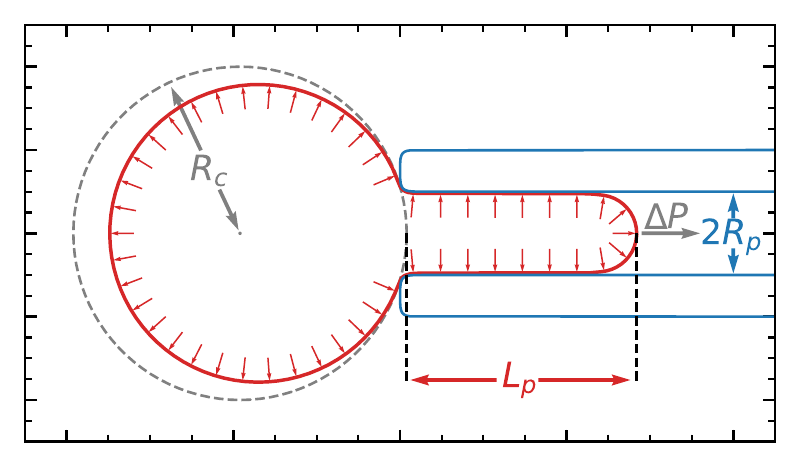}
\caption{\label{fig:ParDescriptionSingle} Cell model under the micropipette aspiration simulation.  Grey dashed line indicates cell membrane initial condition; red solid line indicates cell membrane after pressure application. The red arrows inside the membrane designate the internal pressure of the cell. Micropipette walls are depicted in blue and  pressure is indicated by a grey arrow. Radius of the cell at the initial condition, micropipette width, deformation length and aspiration pressure are defined as $R_c$, $2R_p$, $L_p$ and $\Delta P$, respectively.}
\end{figure}

For our simulations, we use an initially circular cell with radius $R_c = Nd_0/(2\pi)$, and a target area $A_0=\pi R_c^2$ in a membrane composed of $N$ active particles. So, in the absence of external pressure or interaction with the micropipette, the system is in mechanical equilibrium. The channel is represented by a cavity of width $2R_p=0.5R_c$, being $R_p$ the cavity radius. Along with the simulation, the membrane region inside the cavity is aspirated by a pressure $\Delta P$.  A grey dashed line represents the initial membrane configuration in Figure \ref{fig:ParDescriptionSingle}. The blue color indicates the micropipette channel, and solid red lines represent cell membrane and deformation length, $L_p$, after aspiration. The red arrows inside the membrane's final condition designate the internal pressure provided by the volume potential. 

Our simulation is two-dimensional, so force by unit length is our pressure measure.
We prepare the simulation by adjusting the cell to contact the micropipette channel and setting aspiration pressure to zero, making the initial simulation condition similar to a typical micropipette aspiration experiment~\citep[e.g.][]{Hochmuth2000}.  Purely linear forces control particle-wall interactions in the simulations.

\section{\label{sec:results}Results}

In the following subsections we will present several simulations focused in the map of the physical properties of the proposed model, such as the cell compressibility, the bending rigidity influence,  the activity, relaxation time and elasticity.
To simplify the connection between our simulations and experiments, we rescale all physical unities with three basic model parameters. For time, we use $\tau$, for mass, we use $1/(\mu\tau)$, and for distance, we use $R_c$, which is a viable parameter to be measured in experiments.

\subsection{Cell Compressibility}

\begin{figure}[htp]
\includegraphics[width=\linewidth]{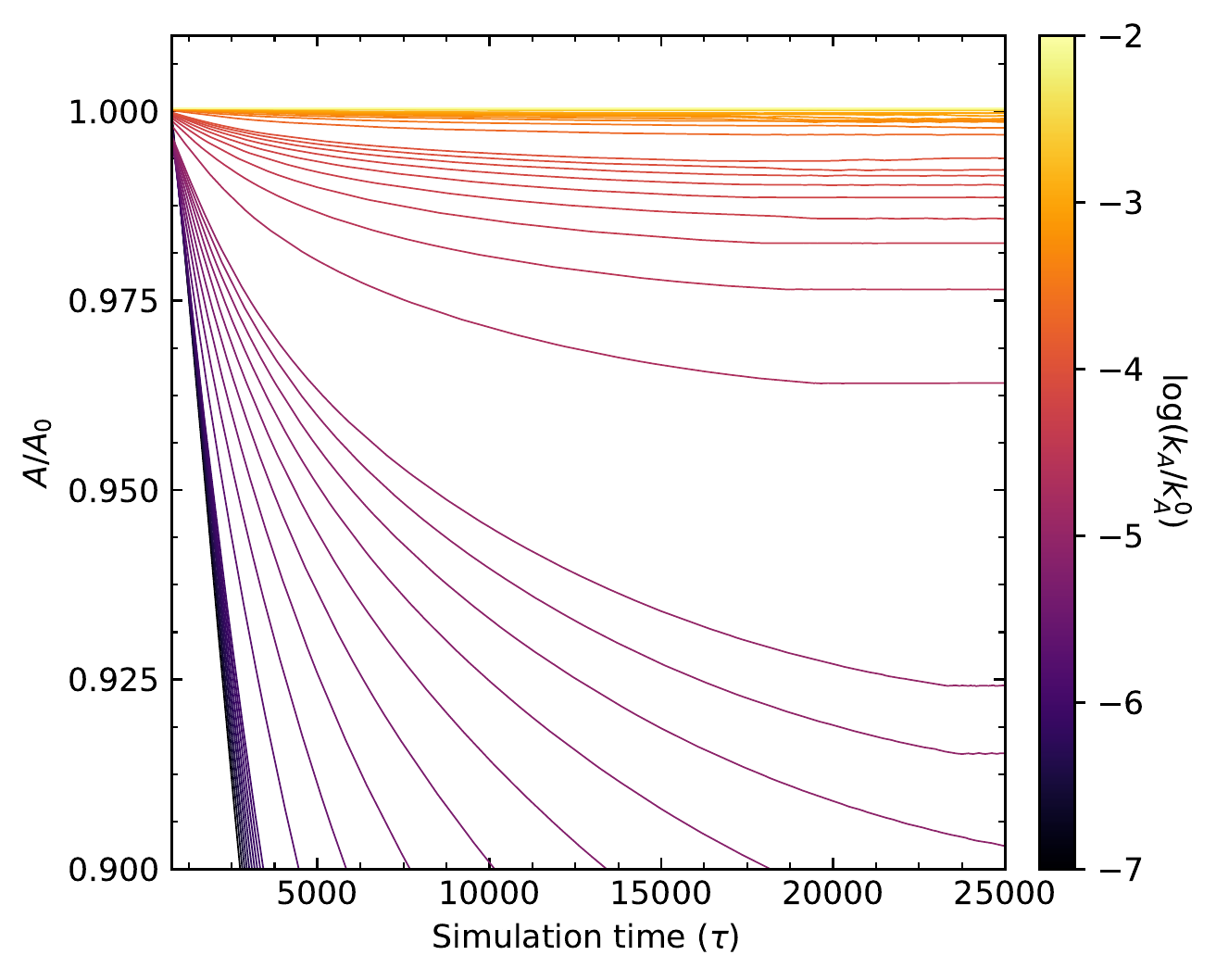}
\caption{\label{fig:EvolkAA0} Evolution of the relative area of the cell. The line colour indicate the value of $\log{(k_A/k^0_A)}$ according to the colour bar, varying from $-7$ to $-2$. For $\log{(k_A/k^0_A)}$ smaller than $-6$ the membrane is totally aspirated. For $\log{(k_A/k^0_A)}$ higher than $-4.15$ the relative volume variation is smaller than $5$ per cent.}
\end{figure}

As mentioned before, the \citet{Teixeira2021} model can be indefinitely compressed. Unless for exceptional cases where the cell can not be deformed due the bending rigidity, such cell model would be totally aspirated inside the micropipette during the experiment. 
We propose the potential presented in Equation \ref{eq:potentialarea} to guarantee a finite compressibility to our model. To verify the effects exclusively from this potential, we fixed the free parameters of our model as $N=500$, $\Delta P=7.5\,(\tau\mu)^{-1}$, $k_b=10^{-8}\, R_c^2/(\tau\mu)$, $k_s=10\,(\tau\mu)^{-1}$ and $v_0=5\times 10^{-6}\,R_c/\tau$. We simulate the aspiration experiment using values of $k_A$ from $10^{-3}\,(\tau\mu R^2_c)^{-1}$ to $10^{2}\,(\tau\mu R^2_c)^{-1}$. The simulations where stopped when the cell reaches the equilibrium state. For sake of simplicity we define $k^0_A = 10^{4}\,(\tau\mu R^2_c)^{-1}$. In Figure \ref{fig:EvolkAA0} we present the evolution of the relative area of the cell, $A/A_0$, for several values of $k_A$.

Figure \ref{fig:EvolkAA0} indicates that for small values of $\log{(k_A/k^0_A)}$ the cell takes long times to evolve and the final area is much smaller than the equilibrium area. Note that all logarithms used in this work are in base $10$. For $\log{(k_A/k^0_A)}$ smaller than $-6$ the cell is totally aspirated. For $\log{(k_A/k^0_A)}$ equals to $-5$, $-4$ and $-3$ the relative area variation is, respectively, $7.58$, $0.63$ and $0.03$ per cent.

\subsection{Bending rigidity investigation} 

\begin{figure}[htp]
\includegraphics[width=\linewidth]{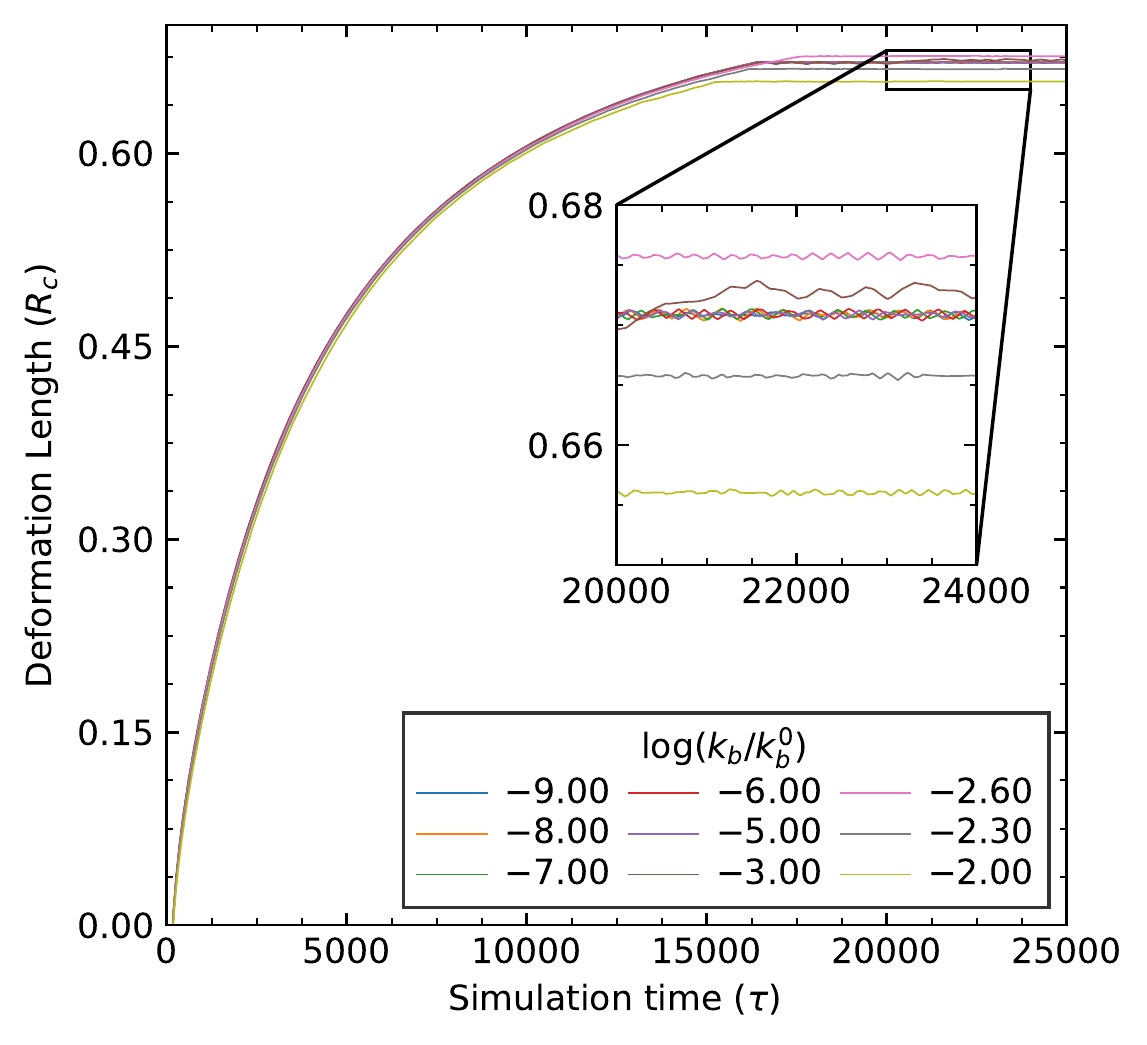}
\caption{\label{fig:Evolkbvar} Evolution of the deformation length during the micropipette aspiration experiment for values of $\log{(k_b/k^0_b)}$ from $-9$ to $-2$. As can be seen in the inset region, the deformation length is exactly the same, under the fluctuations, for $\log{(k_b/k^0_b)}$ smaller than $-5$. The equilibrium value of the deformation length for $\log{(k_b/k^0_b)}$ higher than $-5$ can be higher or smaller than the value for a smaller $\log{(k_b/k^0_b)}$, suggesting that for a higher $k_b$ the bending tension affects the configuration but not the elasticity.}
\end{figure}

To verify the influence of the bending rigidity in our model, we follow the same procedure used to describe the cell compressibility. We fix the free parameters to the same values as mentioned before, letting the parameter $k_b$ vary between $10^{-9}\, R_c^2/(\tau\mu)$ and $10^{-2}\, R_c^2/(\tau\mu)$. For convenience we define $k^0_b = 1\, R_c^2/(\tau\mu)$. In Figure \ref{fig:Evolkbvar} we present the evolution of the deformation length $L_p$ for several values of $k_b$.

In this figure we can notice that the bending rigidity does not play an important role in our model, since the membrane walls that are inside the micropipette do not experiment any bending tension. For $\log{(k_b/k^0_b)}$ smaller than $-5$ the equilibrium point of the evolution is exactly the same. For $\log{(k_b/k^0_b)}$ higher than $-5$ we can observe fluctuations of less than $3$ per cent in the deformation length (see inset in Fig.\ref{fig:Evolkbvar} ). However, since the deformation length decrease does not follow the bending rigidity increase, we do not expect they have a direct correlation.

\subsection{Activity and Relaxation Time}

\begin{figure}[htp]
\includegraphics[width=\linewidth]{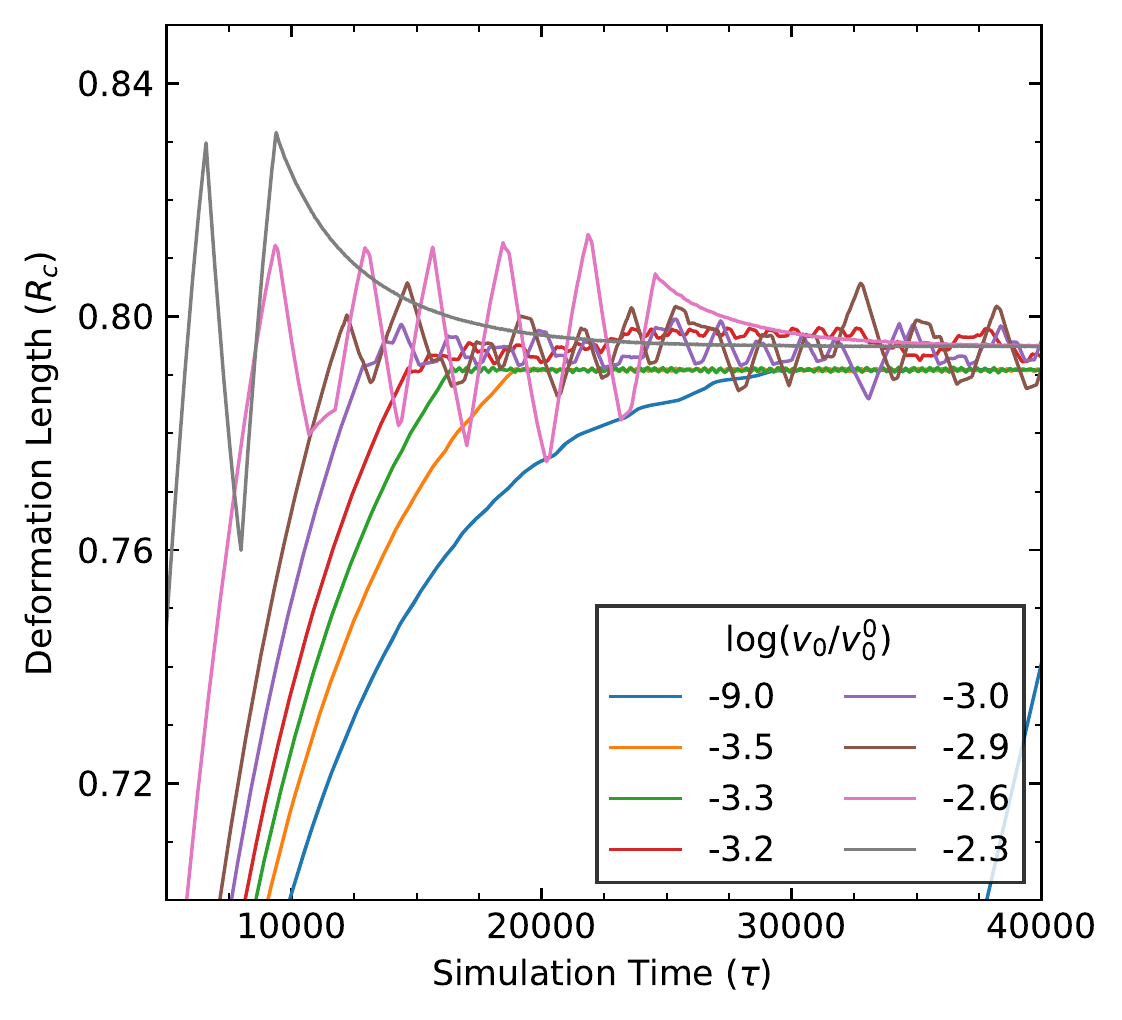}
\caption{\label{fig:v0fac} Evolution of the deformation length for several values of $\log(v_0/v_0^0)$. Three main scenarios can be observed. For $\log{(v_0/v_0^0)}{\leq} -3.3$ the system reaches a well-defined equilibrium state with small fluctuations and a relaxation time towards equilibrium decreasing for higher values of $v_0$. When $-3.3{<}\log{(v_0/v_0^0)}{<}-2.6$ the internal energy leads the system to high fluctuations around the equilibrium. In that case  $\log(v_0/v_0^0){\geq}-2.6$ the system evolves rapidly with highly unstable fluctuations, forcing the system to a rotational state to dissipate the internal energy and slowing down the relaxation time.}
\end{figure}

Cell activity acts as an extra force in the system. During aspiration, the resultant force towards the micropipette orientates the membrane particles speeding up the aspiration process. In Figure \ref{fig:v0fac} we present the evolution of the deformation length for several values of $v_0$, where $v_0^0=10^{-2} R_c/\tau$. The parameters in this simulations are $N=500$, $\Delta P=7.5\,(\tau\mu)^{-1}$, $k_b=10^{-9}\,R_c^2/(\tau\mu)$, $k_s=10\,(\tau\mu)^{-1}$ and $k_A=10^2\,(\tau\mu R^2_c)^{-1}$.

Note in Fig. \ref{fig:v0fac} that for $\log{(v_0/v_0^0)}{<}-3.3$ activity speeds up the relaxation process and leads to an equilibrium state with very small fluctuations. For $\log{(v_0/v_0^0)}$ slightly higher than $-3.3$ activity rules the evolution, leading to an asymptotic state region with large fluctuations.

\begin{figure}[htp]
\includegraphics[width=\linewidth]{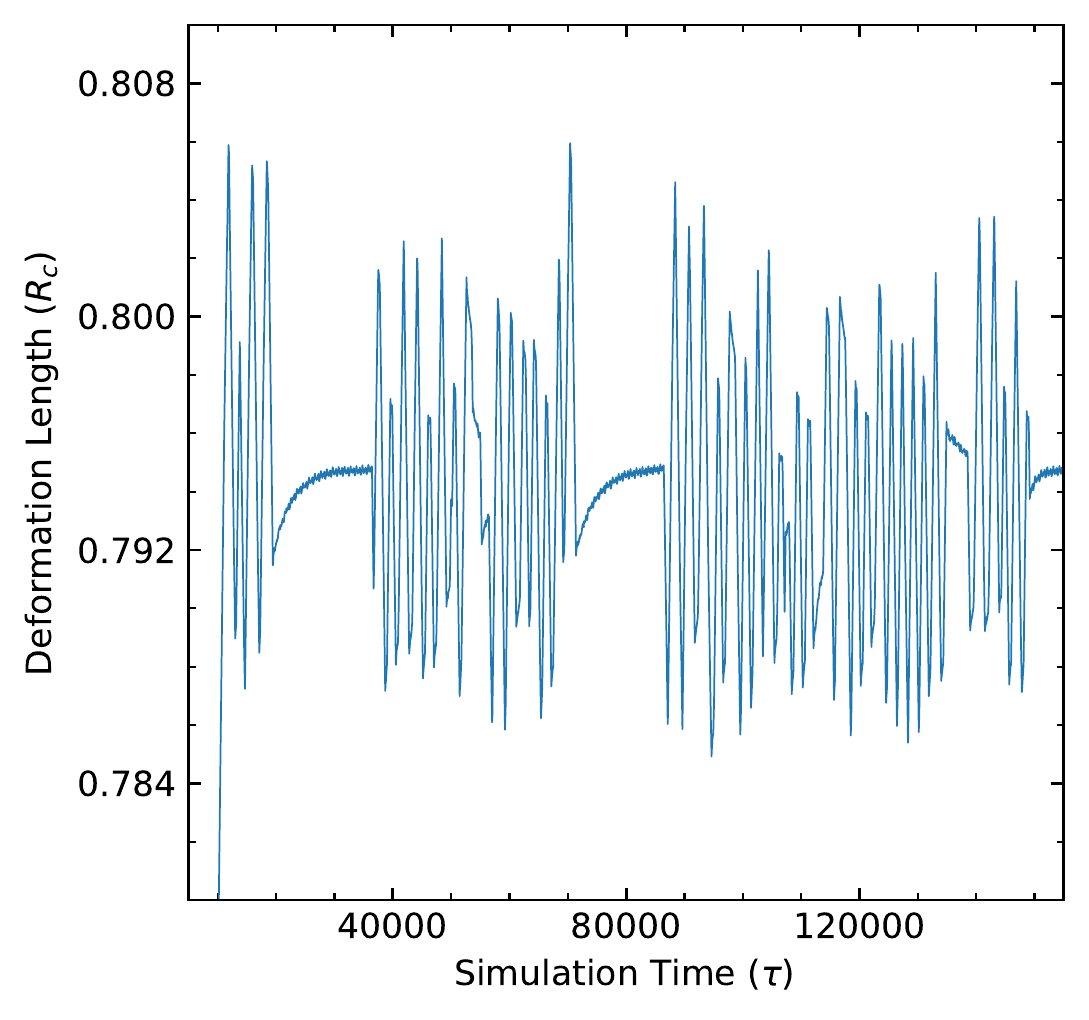}
\caption{\label{fig:v0var} Evolution of the system with $\log{(v_0/v_0^0)}{=}-2.8$. The system evolves rapidly to the maximum deformation length with high fluctuations. In this configuration the final configuration is not well-defined, system particles change from  rotational alignment to direct alignment with the aspiration force alignment.}
\end{figure}

For $\log{(v_0/v_0^0)}$ higher than $-2.6$  fluctuations from the activity are high enough to force the system to an alternative configuration to dissipate the internal energy. This configuration is the rotational state already described by \citet{Teixeira2021}, where particles in the membrane orientate towards the next neighbour and not, as previously, in the resultant force direction.
For example, Figure \ref{fig:v0var} shows the system evolution for $\log{(v_0/v_0^0)}{=}-2.8$. In this situation, the rotational configuration dominates, leading the system to a non-stable, rotating configuration.

\subsection{Elasticity}

The elasticity of a membrane composed by several springs connected can be described by $\rho_s = d_0k_s$. Since the distance between the particles in the membrane scales with the total number of particles connected, $\rho_s$ is valid even for continuum problems.

From the  forces balance, we obtain the value of $\rho_s$ for a two-dimensional membrane under micropipette aspiration,
\begin{equation}\label{eq:rho_s_exact}
    \rho_s = \frac{2\pi R_c R_p \Delta P}{l'-l}
\end{equation}
where $l$ is the cell initial perimeter and $l'$ is the cell perimeter after the aspiration.

Notice that, if we assume the deformation length is much smaller than the total cell perimeter, similar to the assumption made by \citet{Theret1988}, we can approach $(l'-l)$ to $2L_p$. The result of this approach can be seen in equation \ref{eq:rho_s_approach}, a more convenient equation for experimental comparison.

\begin{equation}\label{eq:rho_s_approach}
    \rho_s = \frac{\pi R_c R_p \Delta P}{L_p}
\end{equation}

\begin{figure}[htp]
\includegraphics[width=\linewidth]{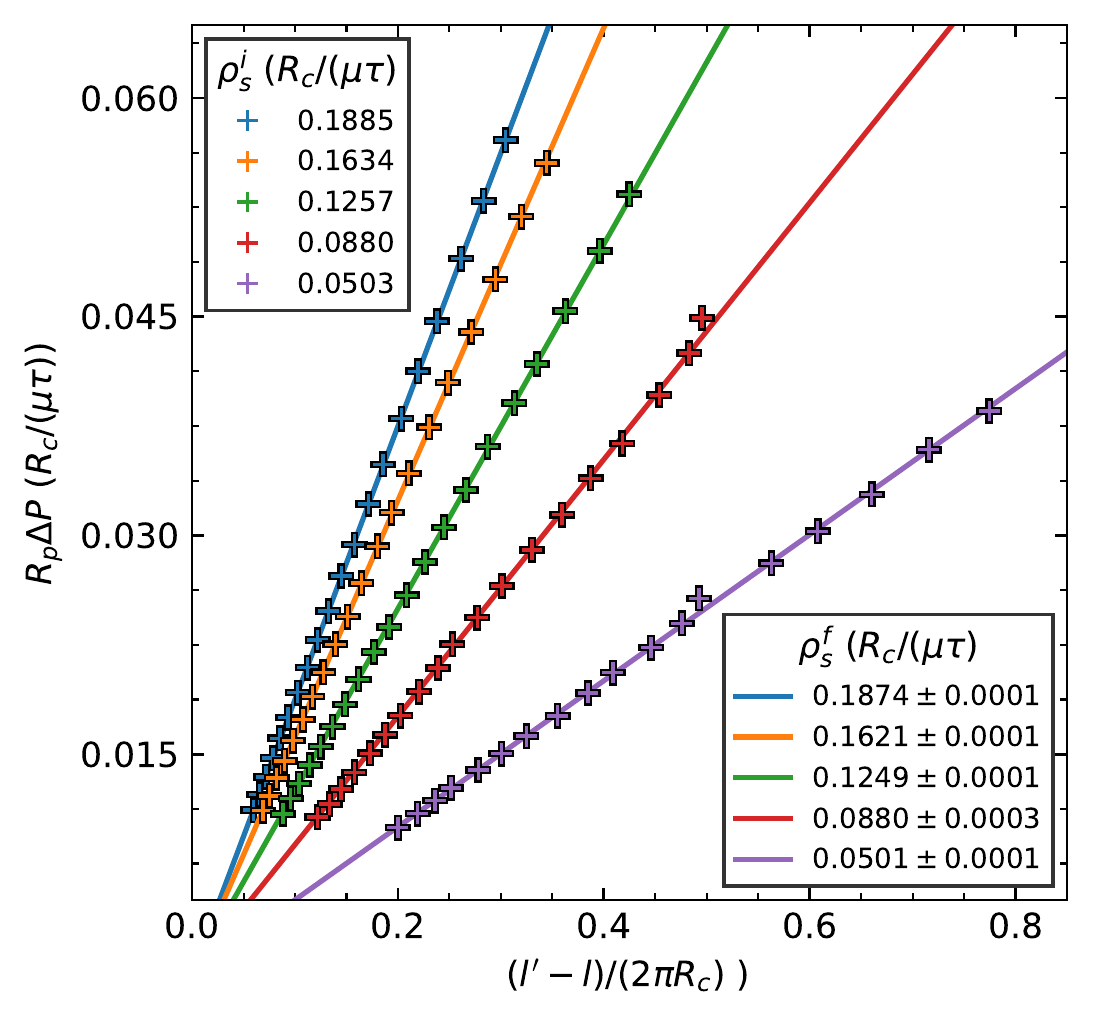}
\caption{\label{fig:rho_s} Relation between between $R_p \Delta P$ and $2\pi R_c/(l'-l)$. The data points of the same colour are the result of simulations performed with different aspiration pressure. The colours indicates the input $\rho_s$, $rho_s^i$. The solid lines are the linear fits, which provides a measure for the $\rho_s$ parameter, in this case defined as $\rho_s^f$.}
\end{figure}

We validate the relation between the experimental measurable parameters $\Delta P$, $R_p$, $R_c$, $l'$ and $l$ by performing several simulations varying the input $\rho_s^i$, and calculating the resulting $\rho_s^f$ by linearly fitting  $R_p \Delta P$ $\times$ $2\pi R_c/(l'-l)$. Figure \ref{fig:rho_s} presents simulations performed for five distinct values of $\rho_s^i$. The data points of the same colour in this figure represents distinct aspiration pressures for the same $\rho_s^i$. The solid lined are the linear fits, which provides the value of $\rho_s^f$.

\section{\label{sec:conclusion}Conclusion}

In this work we use a simple active particle ring model to reproduce numerically an aspiration experiment with cells preserving area and perimeter.

Our simulations indicate that the cell internal pressure  is an essential parameter for the description of  cell dynamics under the micropipette experiment. When simulated without the area conservation potential or with the area conservation potential constant smaller than $10^{-6}\,(\tau\mu R^2_c)^{-1}$, the cell would always be totally aspirated inside the micropipette. For area potential constant between $10^{-5}\,(\tau\mu R^2_c)^{-1}$ and $10^{-4}\,(\tau\mu R^2_c)^{-1}$ cell area varies less than $10$ per cent, and for values higher than $10^{-4}\,(\tau\mu R^2_c)^{-1}$, cell area varies less than $1$ per cent.

Our simulations indicate that the bending rigidity has little effect in the model's elastic properties. Varying the bending rigidity more than seven orders of magnitude  produces fluctuations of less than $3$ per cent in the deformation length. Moreover the fluctuations do not follow the  increase of the bending rigidity, indicating that higher bending tension just traps the system in some metastable equilibrium point.
We notice that the cell internal pressure, which keeps the closest possible to the equilibrium volume, and the membrane springs, which keep the perimeter closest to its equilibrium value, naturally drive the cell to a round-shaped configuration. 

Particles' activity
 in the membrane acts like an extra force. At low values, the activity aligns  particles speeding up the evolution of the system  to the stationary state. At some limits, activity may drive the system to unstable configurations. The three activity regimes observed in our simulations are in agreement with the regimes described by \citet{Teixeira2021}.

Finally, we use the forces equilibrium to relate  the spring's constant and spring's size with parameters macroscopically measured in the experiment.  That is, we use an equation to measure the value of $\rho_s$ using only parameters that can be measured in experiments, such as, $L_p$, $R_c$, $R_p$ and $\Delta P$. To validate our simulations, we compare our input $\rho_s$ with the obtained from the measured using the simulation data, which resulted in less than $1$ per cent of difference. This numerical experiment shows that our simulation is capable of describing  an elastic membrane and the microscopic parameters of the membrane can be recovered. 

Due to its simplicity,  the model presented here can be easily used to simulate cell aggregates under the micropipette aspiration experiments. 

\section*{Acknowledgements}
G. O. and E.F.T.  thanks the Brazilian funding agencies CNPq and Capes. L.G.B. acknowledges the Max-Planck Institute of Ploen, where part of this work was developed. The simulations were performed on the IF-UFRGS computing cluster infrastructure.

\bibliographystyle{apsrev4-2}
\bibliography{apssamp}

\end{document}